\documentclass[useAMS,usenatbib]{mn2e}

\usepackage{times}
\usepackage{txfonts}
\usepackage{bm}

\usepackage[pdftex]{color}
\usepackage[pdftex]{graphicx}
\usepackage[pdftex]{hyperref}

\newcommand{\apj}{ApJ}
\newcommand{\aj}{AJ}

\newcommand{\sm}{\rm M_\odot}
\newcommand{\dlens}{D_{\rm L}}
\newcommand{\dsource}{D_{\rm S}}
\newcommand{\dls}{D_{\rm S}-D_{\rm L}}

\newcommand{\vt}{{\rm v}_t}

\begin{document}

\title{Using microlensed quasars to probe the structure of the Milky Way}

\author[Wang \& Smith]{Jian Wang$^1$ and
  Martin C. Smith$^{2,3}$\thanks{msmith@kiaa.pku.edu.cn}
\medskip
\\$^1$Department of Physics and Tsinghua Center for
Astrophysics, Tsinghua University, Beijing 100084, China
\\$^2$Kavli Institute for Astronomy and Astrophysics, Peking
University, Beijing 100871, China
\\$^3$National Astronomical Observatories, Chinese Academy of Sciences, Beijing 100012, China}

\date{Accepted ........Received .......;in original form ......}

\pubyear{2010}

\maketitle

\begin{abstract}
This paper presents an investigation into the gravitational
microlensing of quasars by stars and stellar remnants in the Milky
Way. We present predictions for the  all-sky microlensing optical
depth, time-scale distributions and event rates for future large-area
sky surveys. As expected, the total event rate increases rapidly with
increasing magnitude limit, reflecting the fact that the
number density of quasars is a steep function of magnitude.
Surveys such as Pan-STARRS and LSST should be able to detect more than
ten events per year, with typical event durations of around one month.
Since microlensing of quasar sources suffers from fewer degeneracies
than lensing of Milky Way sources, they could be used as a powerful
tool for recovering the mass of the lensing object in a robust, often
model-independent, manner.
As a consequence, for a subset of these events it will be possible to
directly `weigh' the star (or stellar remnant) that is causing the
lensing signal, either through higher order microlensing effects and/or
high-precision astrometric observations of the lens star (using, for
example, Gaia or SIM-lite). This means that such events could play a
crucial role in stellar astronomy.
Given the current operational timelines for Pan-STARRS and LSST, by the
end of the decade they could potentially detect up to 100 events. 
Although this is still too few events to place detailed constraints on
Galactic models, consistency checks can be carried out and such
samples could lead to exciting and unexpected discoveries.
\end{abstract}


\section{Introduction}

Gravitational microlensing is now established as a powerful tool for
various aspects of Galactic science. The concept of gravitational
microlensing, where a background object is temporarily brightened by
the gravitational lensing effect of a passing foreground object, is
believed to have been initially conceived by Einstein around 1912
\citep{ren97}.
As he stated in a later publication \citep{ein36}, his opinion was that
"there is no great chance of observing this phenomenon".
Fortunately the idea was later resurrected by several authors
\citep{lie64,ref64}, in particular \citet{pac86}. Following the
proposal of Paczy\'nski, many experiments began to hunt for microlensing
events towards the Galactic bulge and the Magellanic Clouds: OGLE
\citep{uda92}, EROS \citep{aub93}, MACHO \citep{alc93}, MOA
\citep{bon01}.
To date thousands of microlensing events have been detected
toward several directions, including the Galactic bulge
\citep{tho05,sum06,ham06}, Magellanic Clouds
\citep{tis07,wyr09,wyr10}, and M31
\citep{cal05,dej06} and even in surveys of bright nearby stars \citep{fuk07,gau08}.

The purpose of this paper is to deal with a previously under-exploited
set of targets, namely quasars. Although quasar microlensing has been
observed for many years \citep[e.g.][]{cha79,irw89,ang08} this has,
with a few exceptions \citep[e.g.][]{bya70}, almost exclusively been
lensing of background quasars by compact objects in the intervening
(strong) lensing galaxy. In contrast, we are
interested in the lensing of background quasars by stars in the Milky
Way. This is a particularly prescient subject because imminent
large-scale optical surveys, such as Pan-STARRS \citep{kai02} and LSST
\citep{tys02,ive08}, will monitor the whole available sky every
few nights down to magnitudes which are sufficient to monitor several
million background quasars. If large enough samples of such
microlensing events could be accrued then it will be possible to use
them to constrain models of the Galaxy.

One reason why quasars are potentially important targets is that they
break a number of the degeneracies known to inhibit the analysis of
microlensing events from sources within our own Galaxy. In general,
for lensing of Galactic sources it is
difficult to determine the properties of the lensing population
because all of the information about the event is contained in one
observable, namely the event time-scale. As a consequence, there are a
number of degeneracies which need to be broken in order to determine,
for example, the lens mass or distance. For sources
within the Milky Way, in general, we do not know the distance or
transverse velocity of the star which is lensed. For quasars the
interpretation is made significantly easier because we can assume that
the source is at infinity and that the transverse velocity is
negligible. In this case the lens mass is simply a function of the
(known) time-scale and the (unknown) lens distance and velocity,
\begin{equation} \label{equ1}
M \approx 0.3\rm \sm\left({t_{\rm E} \over 30~ \rm day}\right)^2 \left({\dlens \over
4~ \rm kpc}\right) \left({\mu_t \over 9.4~ \rm mas~yr^{-1}}\right)^2,
\end{equation}
where $\dlens$ and $\mu_t$ are the lens distance and proper motion,
respectively, and $t_{\rm E}$ is the event time-scale. This makes the
interpretation of quasar microlensing events significantly
easier than that of Galactic sources and opens up the possibility of
carrying out model-independent mass determinations using high
precision astrometry from, for example, ESA's Gaia mission \citep{per01}.
Even without a measurement of the lens distance/velocity, it could
also be possible to use higher-order microlensing effects such as the
microlensing parallax \citep{gou92} to break the degeneracy. Clearly
quasar microlensing has the potential to become an important avenue
for probing both stellar and Galactic astrophysics.

In this paper we carry out Monte Carlo simulations to address this
issue, presenting an all-sky prediction of the rate and optical depth
of microlensed quasars. The paper is organised as follows. In Section
\ref{sec:model}, we describe the model and theory. In Section
\ref{sec:results}, we present the main results, dealing with the
time-scale distribution (\ref{sec:tE}), the predictions for 
the optical depth and event rate (\ref{sec:event_rate}) and also the
prospects for determining the lens masses (\ref{sec:lens_mass}). We
present a summary and discussion in Section \ref{sec:conclusion}. 

\section{The Model}
\label{sec:model}

We utilise a Monte Carlo simulation to give an estimate of the
rate and optical depth for microlensed quasar events. The details
of the simulation are described in the following subsections. Our
microlensing model is constructed following \citet{kir94} and is
similar to a number of previous works \citep{woo05, han08}.

\subsection{Stellar distribution} \label{mod-mass-distribution}
In this work we want to obtain an all sky estimate of microlensed
quasars by the Milky Way stars, which requires a model for the stellar
mass distribution. To do this we follow the model of \citet[][hereafter
BT08]{bin08}. Unless otherwise stated all parameters are taken from 
Model \uppercase\expandafter{\romannumeral1} in table 2.3 of BT08 and
the solar radius is taken to be 8 kpc.
It is immediately clear that the stellar halo's contribution
is negligible\footnote{Even for lines-of-sight out of the plane, where
the relative contribution of the halo will be at a maximum, the total
mass is negligible compared to the disc. For example, if we adopt the
model of BT08 for the disc, and the model of \citet{dej10} for the
stellar halo then the halo contributes less than 4 per cent of the
total stellar mass for any given line-of-sight.}. Thus, only bulge and
disc stars are considered in this paper.

The bulge is treated as a bar with its longest axis inclined by about
$20^\circ$ to the line from Sun to the Galactic center
and the stellar mass distribution in the bar is given by
\begin{equation} \label{eq:bar}
\rho_{\rm bar}(R, z)=\rho_{\rm bar,0} \left({s \over a_{\rm
    bar}}\right)^{-\alpha_{\rm bar}} e^{-s^2 / r_{\rm bar}^2},
\end{equation}
where $s= \sqrt{x^2 + (y^2 + z^2)/q^2 }$. The bulge is truncated at
a radius $r_{\rm bar}$. The three dimension Cartesian coordinates
($x$, $y$, and $z$) are shown in Fig.~\ref{fig1}. For the bar
parameters we adopt $a_{\rm bar} = 1~\rm kpc$, $\alpha_{\rm bar}=
1.8$, $q = 0.35$, $r_{\rm bar}=\rm 3~kpc$, and $\rho_{\rm bar,0} =
1.22~\rm \sm/pc^3$. We have set the normalisation of the bar
mass so that the bar has the same mass as the original bulge
($0.518\times10^{10}\rm \sm$). 

The disc is modeled as the sum of two double-exponentials, which
correspond to the thin- and thick-disc components. The Galactic disc
density distribution in cylindrical coordinates is
\begin{equation} \label{eq:disc}
\rho_{\rm disc}(R, z)= \Sigma_{\rm disc} e^{-R/R_{\rm disc}}\left({\alpha_0 \over 2z_0}e^{-|z|/z_0}
+ {\alpha_1 \over 2z_1}e^{-|z|/z_1}\right),
\end{equation}
where $\Sigma_{\rm disc} = 1773.12~\rm \sm/pc^2$ is the central
surface density, $R$ is the radial distance, $R_{\rm disc} = 2~\rm
kpc$ is the disc scale-length (both components are assumed to have
the same scale-length), $z_0 = 0.3~\rm kpc$, and $z_1 = 1.0~\rm kpc$
are the scale-heights for the thin and thick components,
respectively, and $\alpha_0 = 0.9333$ and $\alpha_1 = 0.0667$ are
the respective normalisations. The total mass of the disc in this
model is $5.13\times10^{10}{\rm \sm}$.

\subsection{Lens Mass Function} \label{mod-mass-function}
For the mass function we use the four part power-law distribution model
of \citet{kro02},
\begin{equation} \label{eq:imf}
\xi_{0}(M)\equiv {dN\over dM} = k \cases{0.4687M^{-0.3} & $0.01\leq
M/\sm < 0.08$, \cr 0.0375M^{-1.3} & $0.08\leq M/\sm < 0.5$,
\cr 0.0187M^{-2.2} & $0.5\leq M/\sm < 1.0$, \cr 0.0187M^{-2.7} &
$1.0\leq M/\sm < 120$.\cr}
\end{equation}
The normalisation $k$ is obtained through the integral,
\begin{equation}
\int_{0.01}^{120}M\xi_{0}(M)dM=\rho_{\rm disc,solar},
\end{equation}
where $\rho_{\rm disc,solar}$ is the stellar density in the solar
neighbourhood (which can be derived from equation \ref{eq:disc}). We
find that $k=0.762~\sm^{-1}\:pc^{-3}$. Note that unlike some previous
authors we do not adopt different mass functions for different
populations; as argued by \citet{bas10}, there is no clear
evidence to support strong variations in the initial mass function.

We invoke the following simple approach for dealing with stellar
evolution. The objects with initial masses $0.01\sim 0.08~\sm$
and $0.08\sim 1.0~\sm$ are assumed to become brown dwarfs and
main-sequence stars, respectively. The stars with initial mass
$1.0\sim 8.0~\sm$ are assumed to evolve into $0.6~\sm$ white
dwarfs. The stars with $8.0\sim 40.0~\sm$ are assumed to evolve
into $1.35~\sm$ neutron stars, and more massive stars are
assumed to evolve into $5.0~\sm$ black holes.

We assume that the lens number density is proportional to the
stellar mass density given in equations (\ref{eq:bar}) and
(\ref{eq:disc}) (i.e. $n \propto \rho_{\rm bar}+\rho_{\rm disc}$).
For convenience, we rewrite the total stellar distribution equation
as follows,
\begin{equation}\label{eq:rho}
\rho_{\rm total}=\rho_{\rm bar}+\rho_{\rm disc}=\rho_{\rm disc,solar}\chiup(l,b,\dlens).
\end{equation}
From this we can deduce the lens number density
within a mass range, $M$, $M+dM$ and a velocity range $v$, $v+dv$,
as follows,
\begin{equation} \label{equ17}
n(l,b,\dlens,M,v_l,v_b)=\chiup(l,b,\dlens) \xi_0(M) f(v_l,v_b).
\end{equation}
where $l,~b$ are the galactic coordinates and $\dlens$ is the
distance to the lens, $\xi_0$ is the lens mass function given in
equation (\ref{eq:imf}) and $\chiup(l,b,\dlens)$ is the dimensionless mass
distribution (see equation \ref{eq:rho}).

\subsection{Lens luminosity}
\label{sec:lens_lf}
In addition to the lens mass, we also consider the lens
luminosity. This allows us to investigate the possibility of detecting
the lenses and measuring their proper motion and distance, which will
break the degeneracy in equation (\ref{equ1}) and allow a
model-independent measurement of the lens mass.

For the main-sequence stars we estimate the lens brightness using
the mass-luminosity relation of \citet{cox98}, while we assume that
all non-main-sequence lenses are dark. The mass-luminosity relation of
\citet{cox98} is given in the $V$-band in the Johnson-Cousins $UBVRI$
system. In order to compare with LSST observations (see Section
\ref{sec:opdepth}) we need to transform the $V$-band luminosity into
a luminosity in the LSST $ugrizy$ photometry system. To do this we
use the empirical colour transformations given by \citet{jor06}.

As the lenses are distributed at different distances from the Sun, we
need to carefully consider the Galactic dust extinction.
To do this we utilise the Galactic dust map of \citet[][hereafter
SFD]{sch98}, combining this with a model of the interstellar medium
from BT08, as follows. We first construct a model to calculate the total
extinction along a given line-of-sight for the photometric band $x$,
\begin{equation}
A_x\left({\infty}\right) = k_x \int_0^{\infty}\rho_g dl,
\end{equation}
where $k_x$ is a constant (i.e. this does not vary across the sky) and
the distribution of the interstellar medium ($\rho_g$) is modelled
using equation (2.211) of BT08. We estimate $k_x$ by carrying out a
least-squared fit between $A_x\left({\infty}\right)$ and the
observed value of the total extinction from SFD using $10^6$ lines-of-sight
uniformly distributed across the sky. Once we have estimated $k_x$ we
can calculate the extinction that a lens is subject to using the 
following formula,
\begin{equation}
A_x\left({\dlens}\right) = k_x \int_0^{\dlens}\rho_g dl.
\end{equation}

\subsection{Velocity Distribution} \label{MVD}
To calculate the event rate we need to know the velocities of the
observer, lenses and sources. For simplicity we neglect the motion of
the Earth around the Sun, the so-called parallax effect \citep{gou92},
returning to this briefly in Section \ref{sec:lens_mass}. For the
calculation of microlensing event rate only the relative lens-source
transverse velocity (i.e. the transverse velocity of the lens, in the
lens-plane, relative to the observer-source line of sight),
$\textbf{v}_t$, is needed, which can be written,
\begin{equation}\label{eq:vtrans}
\textbf{v}_t \approx \textbf{v}_{\rm L}-\textbf{v}_O ~~~ {\rm for} ~~\dlens \ll \dsource,
\end{equation}
where $\dlens$ and $\dsource$ are the distances to the lens and source,
respectively, and $\textbf{v}_{\rm O}$ and $\textbf{v}_{\rm L}$ are
the components of the velocity tangential to the line-of-sight for the
observer and lens, respectively\footnote{Note that although our
sources are at cosmological distances, the transverse velocity
equation reduces to the simple geometrical one given in Equation
(\ref{eq:vtrans}) for lenses residing at non-cosmological distances
\citep[see, for example, appendix B of][]{kay86}.}

\begin{figure}
\centering\includegraphics[width=\hsize]{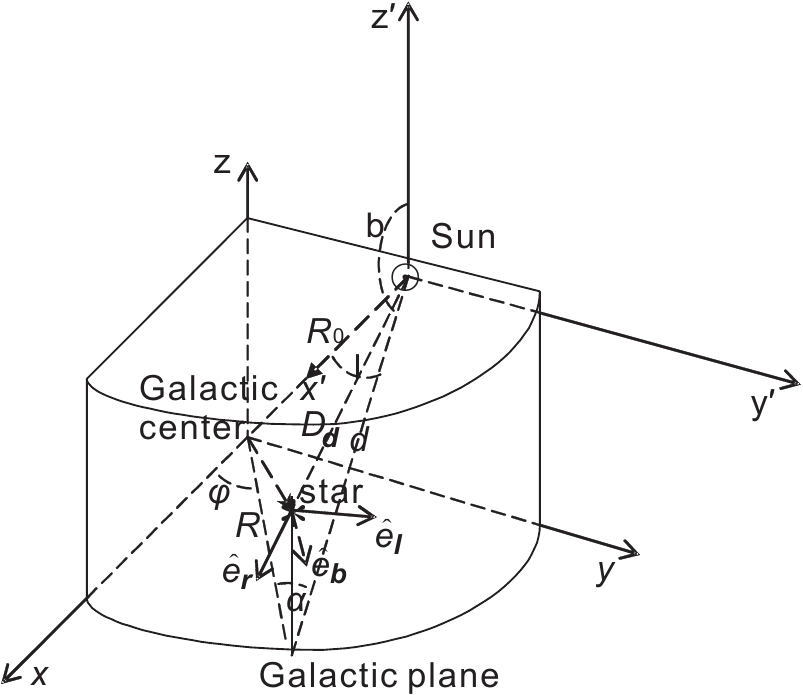}
\caption{Position of a lens star seen from the Sun, where $(\dlens,l, b
)$ are the galactic coordinates of the star, $(R, \varphiup, z)$ are
the Galactocentric cylindrical coordinates, and $R_0$ is the
Galactocentric distance of the Sun. The vectors
$\hat{\textbf{e}}_r$, $\hat{\textbf{e}}_l$, and $\hat{\textbf{e}}_b$
are the unit vectors in the solar-centric spherical
coordinates.}\label{fig1}
\end{figure}

We model the velocity distributions of the lens stars in the Milky
Way using Gaussian distributions, adopting the Solar motion
with respect to the local standard of rest from \citet{sch10}.
We take two separate tri-variate Gaussians for the
thin and thick discs, using the means and dispersions from table 2 of
\citet{smi07}. For the bar stars, the random
velocities are assumed to have Gaussian distributions with
$(\sigma_{x''},\sigma_{y''},\sigma_{z''})=(113.6,~77.4,~66.3)~\rm
km~ \rm s^{-1}$ \citep{han95} along the three axes of the bar.
Here, the coordinates $(x'',~y'',~z'')$ are centred at the
Galactic centre; $x''$ represents the longest axis of the bar, $z''$
is directed towards the north Galactic pole. The bar velocity
dispersion needs to be computed in the Galactocentric cylindrical
coordinate system by,
\begin{eqnarray}
\sigma_{R,\rm bar}^2&=&\sigma_{x}^2\;{\rm cos}^2\varphiup~+~\sigma_{y}^2\;{\rm sin}^2\varphiup,\\
~\sigma_{\varphiup,\rm bar}^2&=&\sigma_{x}^2\;{\rm sin}^2\varphiup~+~\sigma_{y}^2\;{\rm cos}^2\varphiup,\\
\sigma_{z,\rm bar}^2&=&\sigma_{z}^2,
\end{eqnarray}
where
$(\sigma_{x},\sigma_{y},\sigma_{z})=(110.0,~82.5,~66.3)~\rm km~ \rm
s^{-1}$ are the velocity dispersions along the $x,~y,~z$ axis as
shown in Fig.~\ref{fig1}, which are calculated from
$(\sigma_{x''},\sigma_{y''},\sigma_{z''})$ given the assumed bar angle
of $20^\circ$. The resultant velocity dispersion is a
function of the azimuth $\varphiup$. The rotation component of the
bar velocities $v_{\varphiup,\rm bar}$ are estimated following \citet{han95},
\begin{equation} \label{equ12}
v_{\varphiup,\rm bar} = \cases{v_{max}({R \over 1~\rm kpc}) &
$R~<~1~\rm kpc, ~~ solid ~body ~rotation$, \cr v_{max} &
$R~\geq~1~\rm kpc, ~~ flat~ rotation$. \cr}
\end{equation}
where $v_{max}~=~100~\rm km~s^{-1}$ is adopted.

For the calculation of the relative lens-source transverse velocity,
$\textbf{v}_t$, we need to convert the lens velocity components from
Galactocentric cylindrical coordinates into solar-centric spherical
coordinates. The conversion is done as follows,
\begin{equation}\label{equ13}
\left( {\begin{array}{c} v_r \\ v_l \\ v_b
\end{array}} \right)
= \left({\begin{array}{rrr}
 \sin b \cos \alpha & -\sin b \sin \alpha & \cos b \\
 \sin \alpha & \cos \alpha & 0 \\
 \cos b \cos \alpha & -\cos b \sin \alpha & -\sin b
\end{array}}\right)
\left( {\begin{array}{c} v_R \\ v_\varphiup \\ v_z
\end{array}} \right),
\end{equation}
where $(v_r,v_l,v_b)$ are the velocity components in solar-centric
spherical coordinates corresponding to the radial, galactic longitude
and galactic latitude components, respectively. The angle $\alpha$ is
the angle between the lines connecting the Sun, the projected position
of the lens on the Galactic plane, and the Galactic centre (see
Fig.~\ref{fig1}),which can be computed by,
\begin{equation}\label{equ14}
\alpha=\sin^{-1}{R_0~\sin l / R},~~~~R=\sqrt{R_0^2+d^2 - 2R_0~d~\cos
l},
\end{equation}
where $d=\dlens~\sin b$.

\subsection{Quasar Luminosity Function}
\label{sec:qso_lf}
Quasars are very energetic and distant galaxies with an active
galactic nucleus. By assuming an isotropic distribution, the
distribution of quasars can be given by the so called quasar
luminosity function (QLF). A lot of work has been carried out on the
QLF in recent years \citep{boy00, ric06, fon07, hen07, ric09, cro09},
especially with the advent of surveys such as the Two Degree Field
(2dF) and Sloan Digital Sky Survey (SDSS).

\begin{figure}
\centering\includegraphics[width=\hsize]{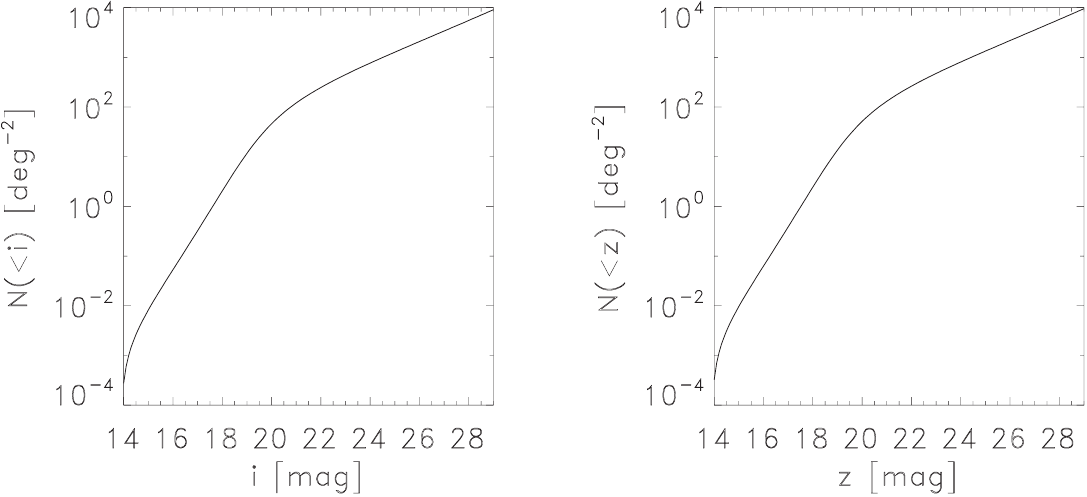}
\caption{The surface number density of quasars as a function of
limiting apparent magnitude in $i$- and $z$-bands.} \label{fig2}
\end{figure}

In this paper we assume that the quasars have a constant surface
number density and follow the method of \citet{lig07} to
give an estimate of this number density. We utilise the $\Lambda$ cold
dark matter ($\Lambda$ CDM) model in a flat universe
(i.e. $\Omega_{m,0}+\Omega_{\Lambda,0}=1$) with $\Omega_{m,0}=0.3$,
$\Omega_{\Lambda,0}=0.7$ and $h=0.7$, where $h$ is the Hubble constant
in units of $100~\rm km~s^{-1}~Mpc^{-1}$. We then simply integrate
the QLF in the redshift interval $z<8$ to derive the surface number
density of quasars in the $i$- and $z$-bands\footnote{Note that
quasars with $z>8$ will have a negligible contribution to the total
surface number density \citep[see fig. 1 of][]{lig07}.}.
We denote the total surface density of quasars down to a
particular limiting magnitude as $N(<i)$ and $N(<z)$ for the two
photometric bands. Fig.~\ref{fig2} shows the resulting distributions.

Since quasars are extragalactic sources, some of the light emitted
from them will be absorbed by the dust in the Milky Way. We take this
into account using the extinction map of SFD.
The motivation for including the $z$-band in our analysis is to see
whether the reduced extinction for this longer wavelength increases
the quasar microlensing event rate compared to $i$-band.

\subsection{Optical Depth and Event Rate}
\label{sec:opdepth}
The microlensing optical depth is defined as the probability that a
particular background source falls into the Einstein radius of any
foreground lens star. As has been mentioned above in Section
\ref{sec:qso_lf}, we assume that the quasars are distributed uniformly
across the sky. If we denote the surface number density as
$n_s(l,~b,~\dsource)$, the optical depth along a given line-of-sight
can be calculated following \citet{kir94}, 
\begin{eqnarray} \label{equ15}
\tau(l,b) &=& {4\pi G \over c^2} \int_0^\infty
n_s(l,b,\dsource)\dsource^2d\dsource
\int_0^{\dsource}d\dlens\chiup(l,b,\dsource)D
\nonumber \\
&& \cdot \int_{M_1}^{M_2} M \xi_0(M)dM \cdot [\int_0^\infty n_s(l,b,\dsource)\dsource^2d\dsource]^{-1}\nonumber \\
&\approx& {4\pi G \over
c^2}\int_0^{\dsource}d\dlens\chiup(l,b,\dsource)D\int_{M_1}^{M_2} M
\xi_0(M)dM,
\end{eqnarray}
where $G$ is gravitational constant, $D\equiv
\dlens(\dls)/\dsource\approx\dlens$ for $\dlens\ll\dsource$, and $M_1$
and $M_2$ are the mass limits of the lens stars. Note that the optical
depth is independent of $n_s(l,~b,~\dsource)$.

The event rate is the number of microlensing events that occur per
unit time, which we again calculate following \citet{kir94},
\begin{eqnarray} \label{equ16}
\Gamma (l,b) &=&{4\sqrt{G} \over c} \int_0^\infty
d\dsource n_s(l,b,\dsource)\dsource^2 \int_0^{\dsource}
d\dlens\chiup(l,b,\dlens)D^{1/2}
\nonumber \\
&& \cdot \int_{M_1}^{M_2}dM \xi_0(M)M^{1/2} \iint dv_l dv_b v_t
f(v_l,v_b)
\nonumber \\
&& \cdot  [\int_0^\infty
d\dsource n_s(l,b,\dsource)\dsource^2]^{-1} \nonumber \\
&\approx& {4\sqrt{G} \over c}\int_0^{\dsource} d\dlens\chiup(l,b,\dlens)D^{1/2}
\int_{M_1}^{M_2}dM \xi_0(M)M^{1/2}
\nonumber \\
&& \cdot \iint dv_l dv_b v_t f(v_l,v_b).
\end{eqnarray}

The time-scale of a microlensing event, $t_E$, which is defined to be
the time for a source to cross the Einstein radius of the lens
\citep[e.g.][]{mao08}, is given by,
\begin{equation} \label{equ18}
t_E~=~{r_E \over \vt}
\approx~ \frac{1}{\vt}[{4GM \over c^2}\dlens]^{1/2} ~~~ {\rm for} ~~\dlens \ll \dsource.
\end{equation}
where $r_E$ is the physical size of the Einstein radius in the lens
plane and $\vt$ is the relative transverse velocity of the lens
(equation \ref{eq:vtrans}).

During a time interval $\Delta t$, the total number of expected
microlensing events is trivially,
\begin{equation} \label{equ17}
N_{\rm events}= \epsilon\Gamma_{\rm total} \Delta t,
\end{equation}
where $\Gamma_{\rm total}$ is the total event rate integrated over all
lines-of-sight and $\epsilon$ is the detection efficiency. For simplicity, we
assume $\epsilon \equiv 100$ per cent. This can be justified when one
considers that the cadence of the surveys such as LSST will be of the
order of a few days, which is suitable to detect practically all
events in our time-scale range (see Section
\ref{sec:tE}). We return to this issue in the following section.

\section{Results}
\label{sec:results}

\subsection{Time-Scale distributions}
\label{sec:tE}

\begin{figure*}
\centering\includegraphics[width=13cm]{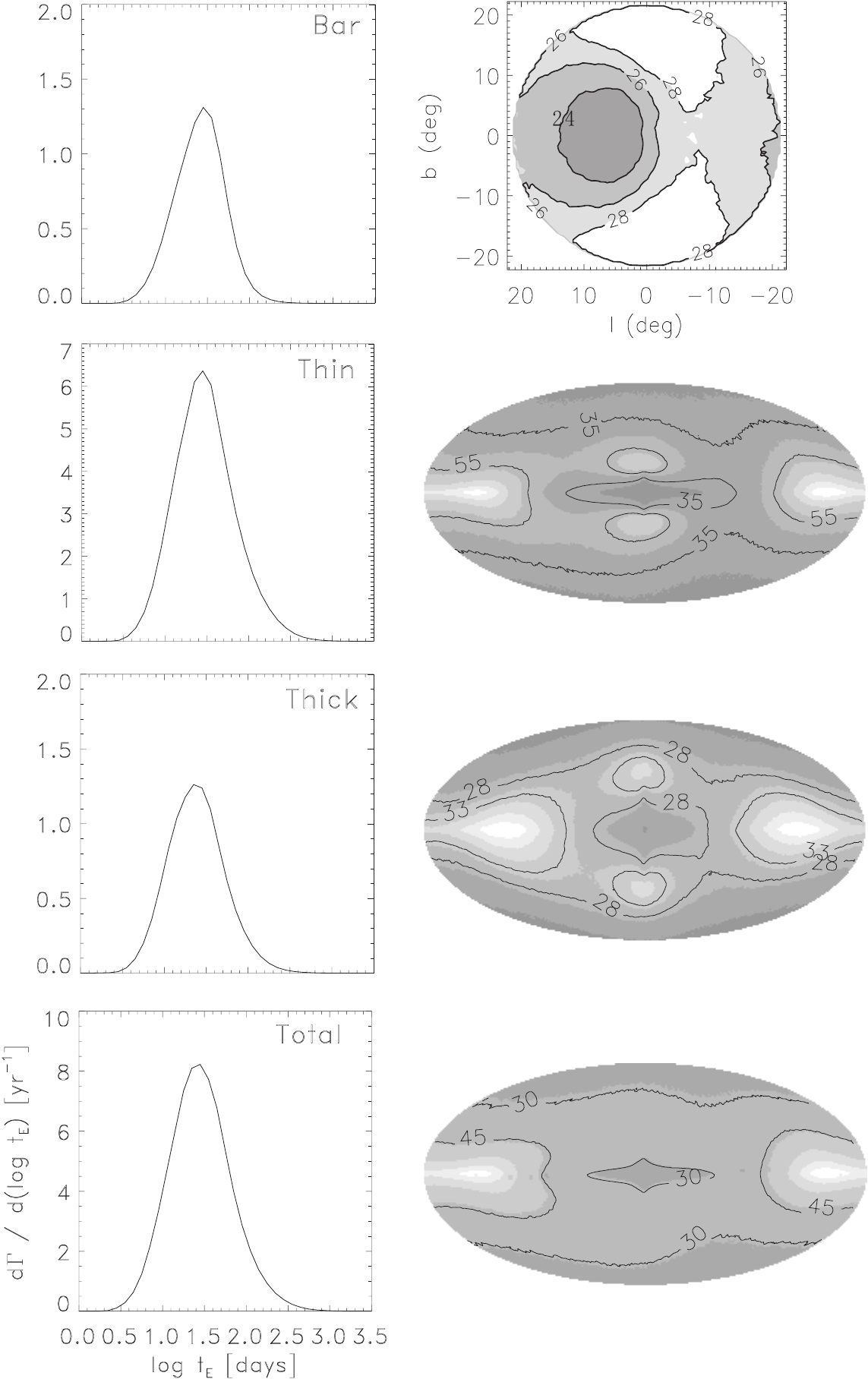}
\caption{The
left-hand panels show the predicted microlensing event rate of
quasars as a function of the time-scale. The right-hand panels show
the corresponding maps of the predicted average event time-scale. We
show the contributions of different Galactic components, the
thin-disc, the thick-disc, the bar, and the combination of bar and
discs from top to bottom, respectively.} \label{fig:tE}
\end{figure*}

Using the model described above, we are able to investigate the
time-scale distributions for quasar microlensing events. Fig.
~\ref{fig:tE} shows the predicted event rate as a function of 
time-scale for the different Galactic components. The average
time-scales for the bar, thin disc, thick disc and total (bar plus
discs) are  
$31.7~\rm days$, $44.4~\rm days$, $33.3~\rm days$ and $41.2~\rm days$,
respectively. The longest time-scales are obtained for thin-disc
events towards $l\approx140$ and $l\approx210$ degrees. In these
regions the mean time-scale can reach as high as 130 days; these
events will be very well sampled and are likely to display microlensing
parallax signatures (see Section \ref{sec:lens_mass}).
Note that the average time-scale for disc-lensing events reaches a
minimum around the central parts of the Galaxy. The average in this
region is reduced due to lenses on the far-side of the bar; these have
large transverse velocities owing to their rotational velocity acting
in the opposite direction to the Sun's. The events with shortest
time-scales are those caused by lenses in the Galactic bar,
because bar lenses have larger velocity dispersions and hence larger
transverse velocities. Furthermore, unlike nearby disc-lenses, they do
not share the disc's rotation velocity.

The combined map in the lower-panel of Fig \ref{fig:tE} closely
resembles the time-scale map for the thin disc, since the majority of
events are caused by thin-disc lenses (see Section \ref{sec:event_rate}).

\subsection{Optical Depth and Event Rate}
\label{sec:event_rate}

Arguably the most important quantity which we can calculate is the
total event rate, i.e. the number of events which we expect future
surveys will be able to detect. This is given in
Table~\ref{tab:event_rate} for various apparent magnitude limits.
Not surprisingly the total event rate increases rapidly with
increasing depth, reflecting the fact that the number density of
quasars rises steeply for fainter limiting magnitudes (as shown in
Fig. \ref{fig2}).

The contributions from the various Milky Way components are shown in
Table~\ref{tab:event_rate}. We can clearly see that the total
event rate is dominated by the disc component, including the thin
and thick disc components. The thick disc component contributes
about $17$ per cent of the whole disc events and therefore plays a
non-negligible role in quasar microlensing. Although the total mass
of the thin disc is $13$ times more massive than the thick disc, the
fact that the thick disc is less concentrated in regions of high
extinction helps to boost its lensing signal. The bar component, with
only around 10 per cent of the total mass of the disc, has
significantly fewer events. As has been stated in
Section \ref{mod-mass-distribution}, the stellar halo will have a
negligible contribution.

\begin{table*}
\begin{tabular}{lcccc}
\hline
\multicolumn{5}{c}{$\Gamma$ ($yr^{-1}$)}\\
$m_{lim}$ (mag) &Bar & Thin Disc &Thick Disc & Total\\
\hline\\
\multicolumn{5}{c}{estimates for $i$-band detection}\\
21   &0.07   &0.55   &0.17   &0.79  \\
23   &0.51   &3.53   &0.86   &4.90  \\
24   &0.97   &6.94   &1.56   &9.54  \\
27   &5.13   &34.75  &6.92   &46.80 \\\\
\multicolumn{5}{c}{estimates for $z$-band detection}\\
21   &0.12   &0.91   &0.23   &1.26  \\
23   &0.77   &5.40   &1.09   &7.26  \\
24   &1.50   &10.31  &1.95   &13.76 \\
27   &7.30   &48.82  &8.26   &64.38 \\
\hline
\end{tabular}
\caption{All-sky microlensing event rates for a range of magnitude
limits. The different columns correspond to events due to the various
Galactic components.}
\label{tab:event_rate}
\end{table*}

\begin{figure*}
\centering\includegraphics[width=\hsize]{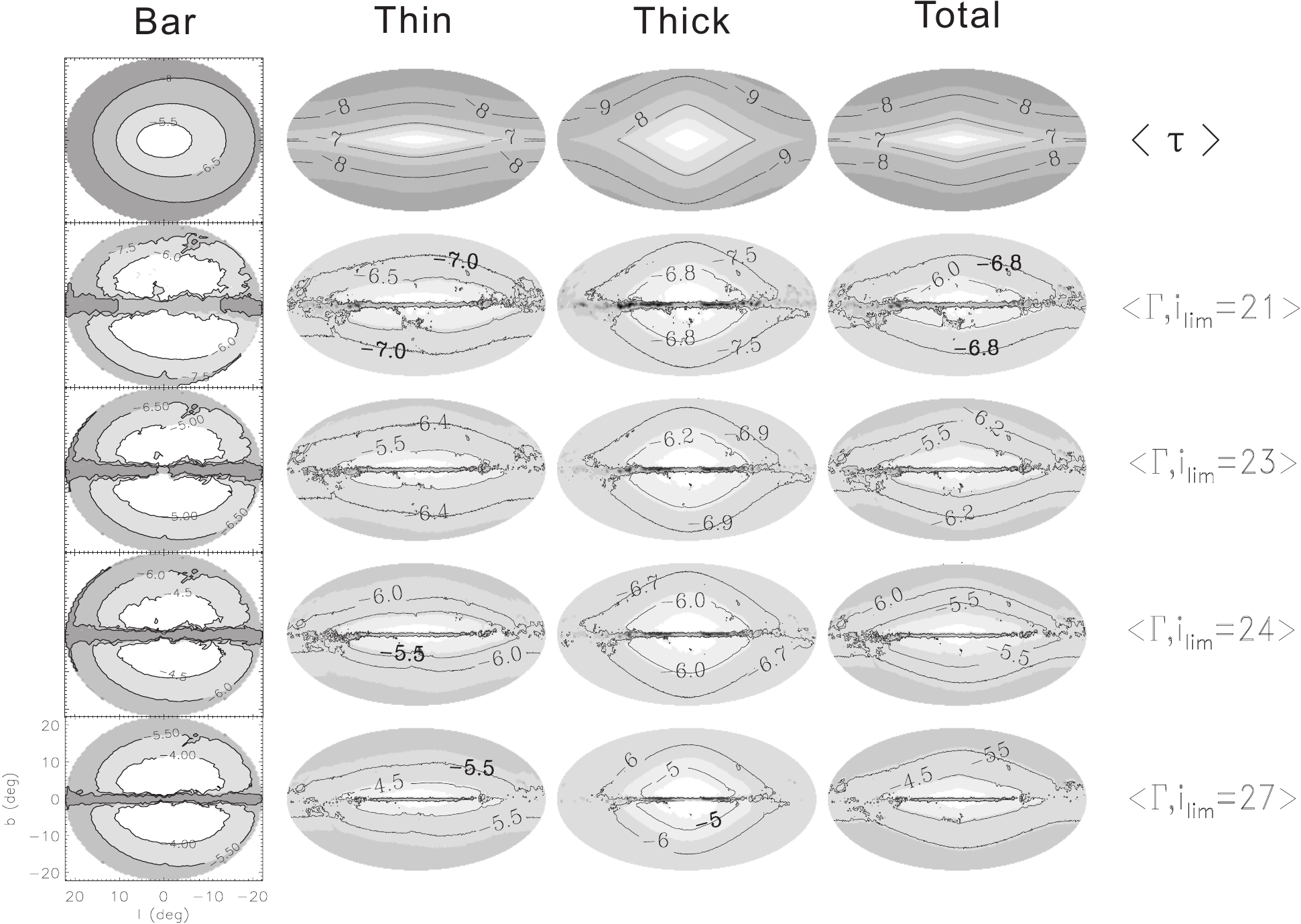}
\caption{Distributions of the microlensing optical depths (top row)
and event rates (bottom four rows). From left to right, the panels in
each column represent maps of the bar component, the thin and
thick disc components and the total contribution from all bar and disc
components. Contours in the top row correspond to the log of the
optical depth, while contours in the bottom four rows correspond to
the log of events per year per square degree for the limiting source
magnitudes shown on the right.}
\label{fig4}
\end{figure*}

The all-sky distributions of average optical depth and the event
rates are presented in Fig.~\ref{fig4}.
The different rows of this figure show the average optical depth
and the event rate for different apparent magnitude limits. As
remarked above, the event rate increases with the increase of the apparent
magnitude limit owing to the increase in source density. The event
rate also increases with decreasing galactic latitude, tracing the
scale-height of the discs. However, as one reaches close to the
Galactic plane extinction becomes important, effectively ruling out
the ability to detect events with $|b| \la 1.5^\circ$. Although
one might expect this problem to be alleviated by operating at longer
wavelengths, the gradient of the extinction (as a function of
$|b|$) is so great that even in the $z$-band the event rate for the
bar and disc only increases by around 30-50 per cent (see Table
\ref{tab:event_rate}).

The most crucial question which we can pose is how many quasar
microlensing events do we expect future surveys to detect? 
There are a number of upcoming surveys which will map the entire
visible sky every few nights. Two important ones are Pan-STARRS
\citep{kai02} and LSST \citep{tys02}, both of which operate with a cadence
of every few nights. We assume a 100 per cent detection efficiency,
which should be a fair approximation since practically all events will
have durations longer than this cadence; from Fig. \ref{fig:tE} we
find that 95 per cent of all events have time-scales greater than 7.5
days (possible limitations to this assumption are discussed in Section
\ref{sec:conclusion}). Pan-STARRS will reach a $i$-band ($z$-band)
magnitude limit of about $i_{lim}~=~22.6$ ($z_{lim}~=~21.5$) mag and
will cover about $30,000$ square degrees, while LSST will reach an
$i$-band ($z$-band) magnitude limit of about $i_{lim}~=~24.0$
($z_{lim}~=~23.3$) mag per visit and cover about $20,000$ square
degrees. In the $i$-band we estimate that Pan-STARRS will detect three
events per year, while LSST will detect around five events per
year. Clearly these numbers depend on the detectability of events,
which is an issue we will return to in the discussion.

As mentioned above, conducting the survey in the $z$-band does not
increase the number of detected events significantly. Furthermore, for
both of these surveys the $z$-band limiting magnitude is around one
magnitude shallower than in the $i$-band, which cancels out any
advantage gained from the lower extinction. As a consequence of these
two factors, the total number of events which would be detected in
this band is only 1 and 4 for Pan-STARRS and LSST, respectively,
meaning that such a survey is optimally carried out in the $i$-band.

In practice we believe that these are lower-limits to the number of
detected events, because the above calculation is based on events for
which the lens and source are aligned to within one Einstein
radius. This corresponds to a minimum amplification of 0.32 magnitudes,
which is easily within reach of these surveys with their milli-magnitude
(or better) photometric precision. For example, if we allow for events
within two Einstein radii (corresponding to a minimum amplification
of around 0.06 mag) then the event rate will increase by a factor of
2. Although it will be challenging to identify events with such small
amplification, it does means that surveys like LSST and Pan-STARRS
could potentially observe around 16 events per year.

\subsection{Lens mass determination}
\label{sec:lens_mass}

It is important to investigate the characteristics of the lens
population, particularly with respect to potential follow-up work on
the lenses.

In Section \ref{mod-mass-function} we described our simplistic
implementation of stellar evolution. From equation (\ref{eq:imf}) it
is trivial to obtain the relative contribution of the different lens
populations. This is summarised in Table \ref{tab:lens_population}. We
see that by far the majority of lenses are main-sequence stars ($\sim
80$ per cent), although a significant fraction are white/brown
dwarfs. Around 1 per cent of events will be caused by neutron stars,
which raises the remote, yet interesting, possibility of quasar lensing from
pulsars. As has been highlighted by previous authors
\citep[e.g.][]{dai10}, pulsar microlensing is of great interest
because for these events it may be possible to determine both the
distance and transverse velocity through radio observations (and
hence, via equation \ref{equ1}, a model independent mass determination).

\begin{figure}
\centering\includegraphics[width=\hsize]{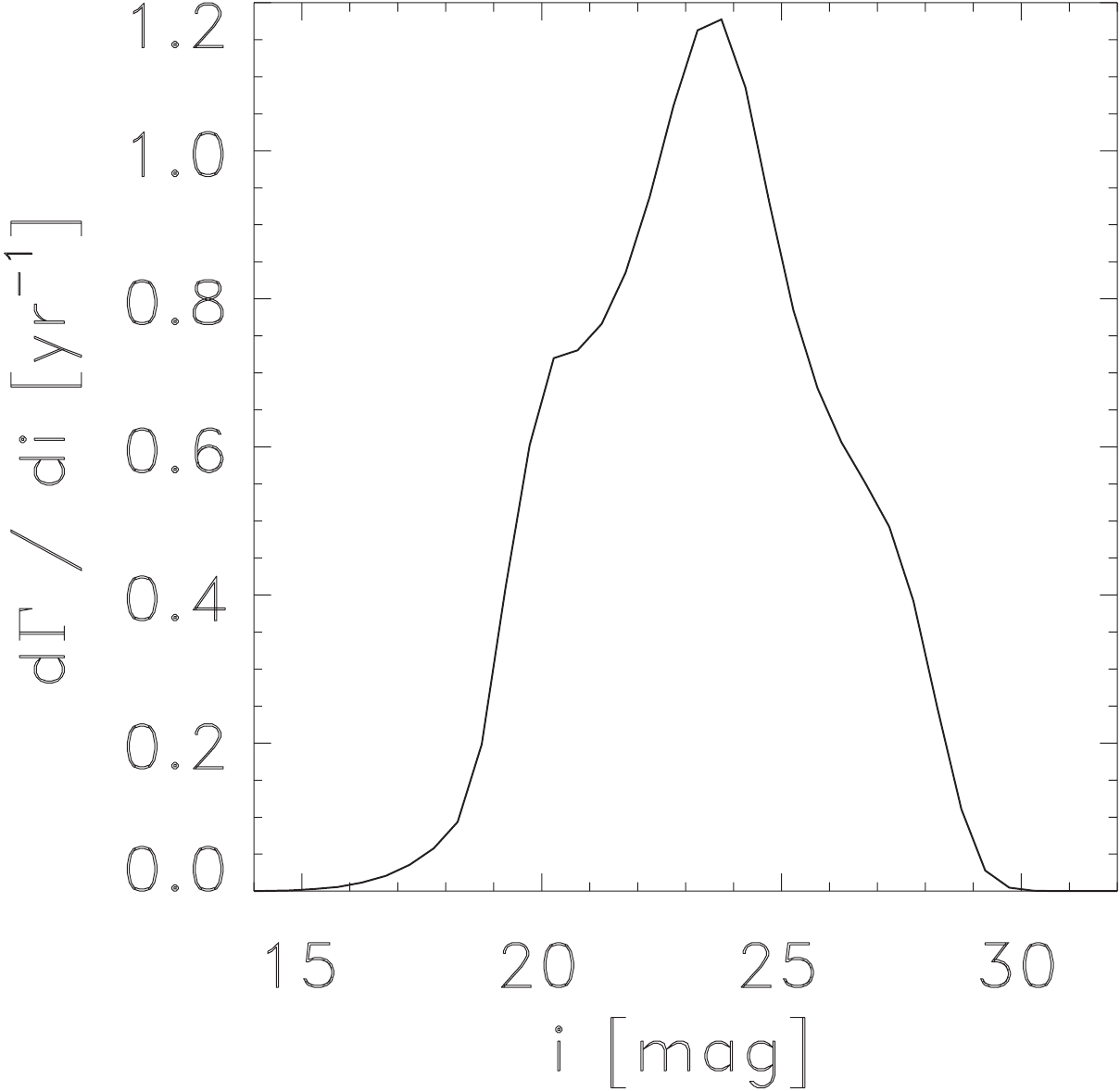}
\caption{The distribution of $i$-band lens magnitudes, after taking
into account extinction.
Note that this corresponds only to the events
caused by main-sequence lenses, i.e. a further 22 per cent of events
have non-main-sequence lenses which we classify as `dark' (see Table
\ref{tab:lens_population}).}
\label{fig:lens_luminosity_marg}
\end{figure}

\begin{figure}
\centering\includegraphics[width=\hsize]{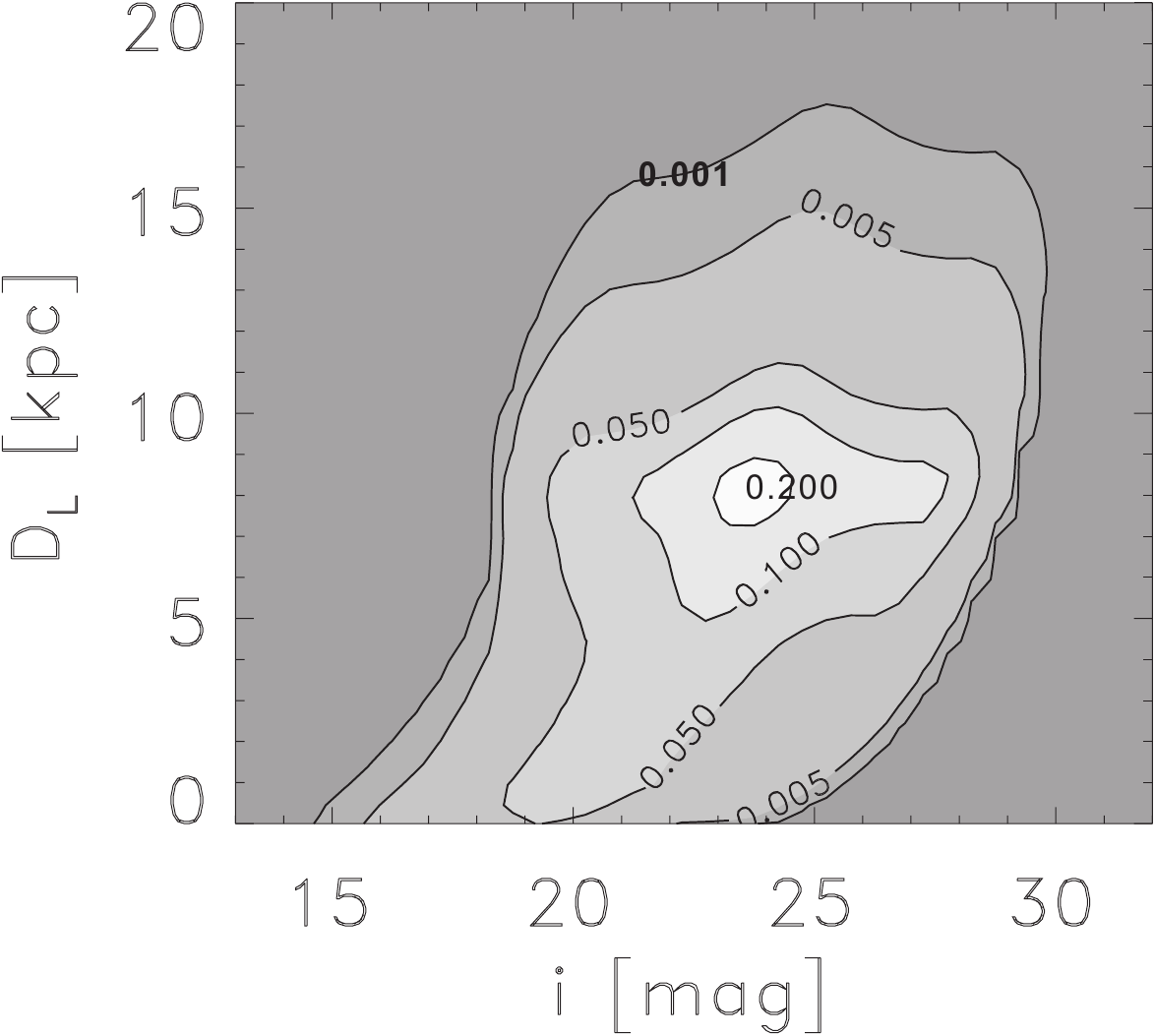}
\caption{The all-sky event-rate map for luminous lenses as a function
of lens distance and magnitude (after incorporating
extinction). Contours correspond to events per year per magnitude per
kpc for a limiting source magnitude of $i=24$.
Note that this corresponds only to the events
caused by main-sequence lenses, i.e. a further 22 per cent of events
have non-main-sequence lenses which we classify as `dark' (see Table
\ref{tab:lens_population}).}
\label{fig:lens_mag}
\end{figure}

\begin{figure}
\centering\includegraphics[width=\hsize]{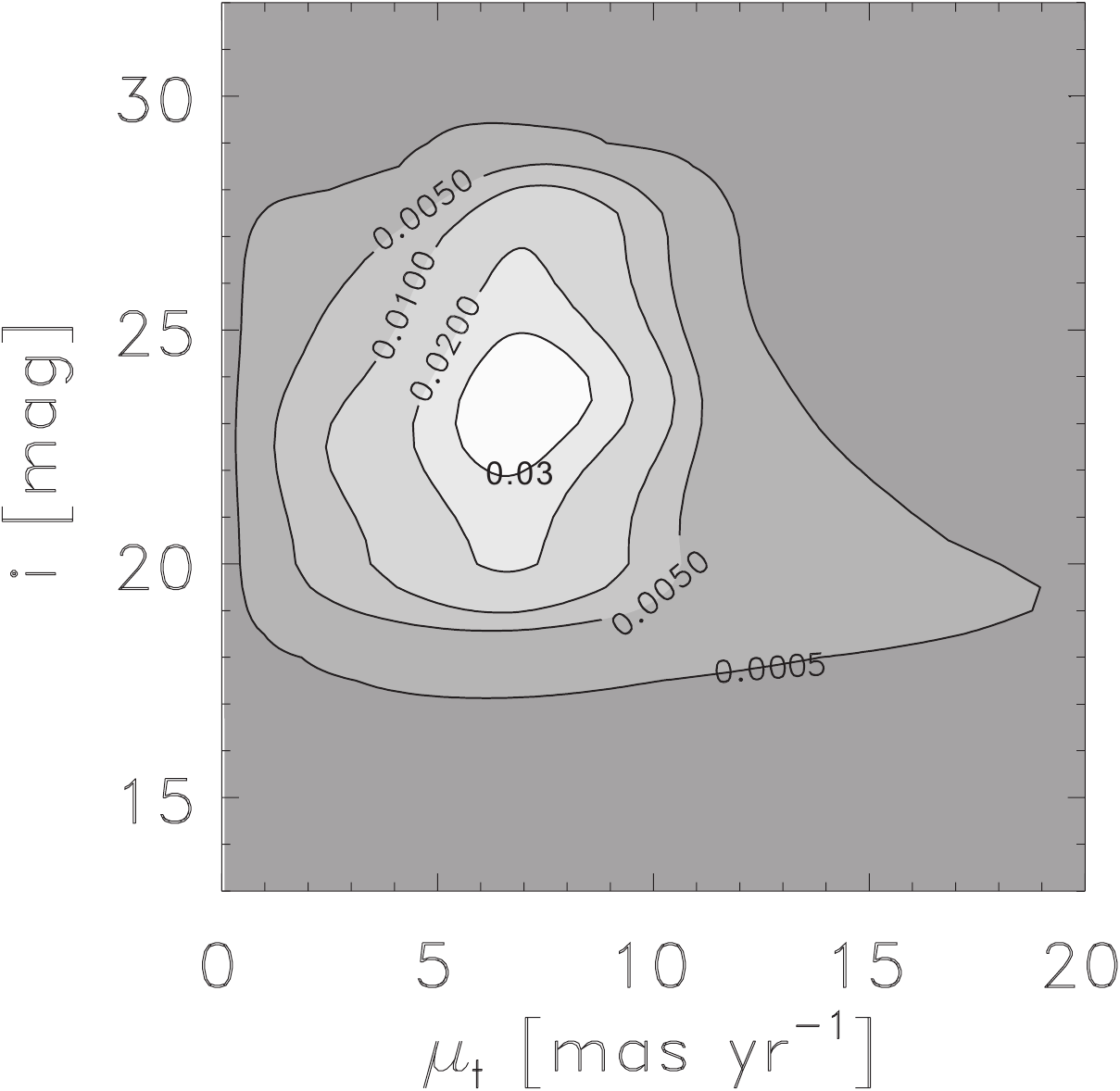}
\caption{Similar to Fig. \ref{fig:lens_mag}, but for lens magnitude
and proper motion.}
\label{fig:lens_propermotion_mag}
\end{figure}

\begin{figure}
\centering\includegraphics[width=\hsize]{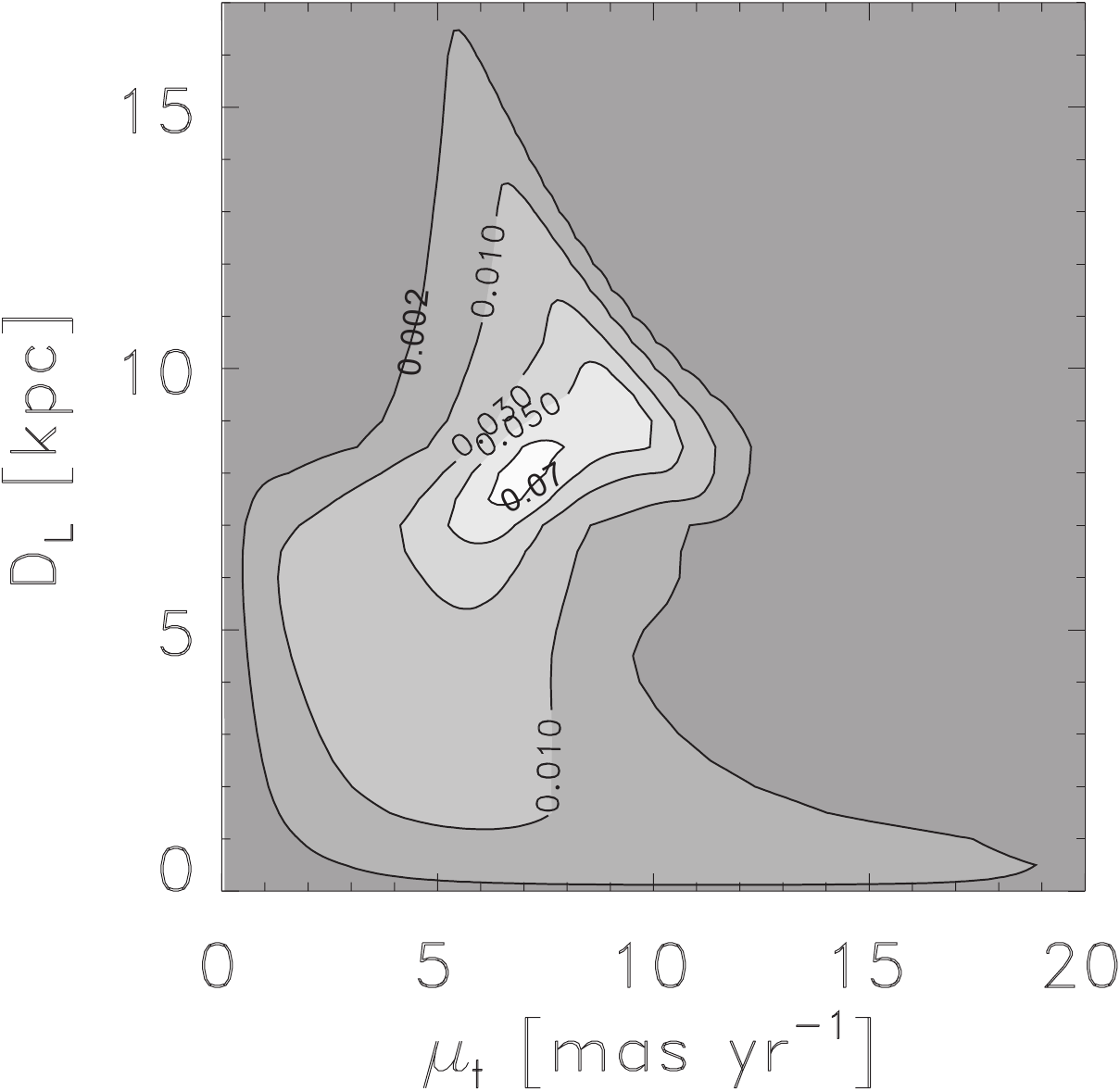}
\caption{Similar to Fig. \ref{fig:lens_mag}, but for lens distance
and proper motion. Note that unlike Figs. \ref{fig:lens_mag} and
\ref{fig:lens_propermotion_mag}, this plot also includes the
non-main-sequence lenses (see Table
\ref{tab:lens_population}).}
\label{fig:lens_propermotion}
\end{figure}

However, the most promising avenue for carrying out model-independent
mass determinations is for main-sequence stars. This is important
because microlensing is the only method which can be applied to single
stars (as opposed to all other methods of mass determination which are
applicable only to stars in binary systems). As has been stated
in the Introduction, to break the microlensing mass degeneracy
requires a measurement of both the distance to the lens and its
transverse velocity.
However, to measure these lens properties
requires the lens to be both bright and nearby.
We investigate this by plotting maps of the event rate for various
lens properties in Figs. \ref{fig:lens_luminosity_marg} to
\ref{fig:lens_propermotion}.

In the near future instruments such as the ESO Gaia mission
\citep{per01} will measure parallaxes and proper motions to
unprecedented accuracy for many millions of relatively bright
stars. Its accuracy degrades rapidly for fainter stars, reaching
accuracies of around 0.1-0.2 mas for stars at the magnitude limit of
around $G = 20$ mag \citep{bai09}. This corresponds to a parallax (and
hence distance) error of around 20 per cent at a distance of 2
kpc. Unfortunately, as can be seen in
Fig. \ref{fig:lens_luminosity_marg}, we predict that very few lenses
will be brighter than 20th magnitude in $i$-band. In
Fig. \ref{fig:lens_mag} we have plotted the map of event rate as a
function of lens brightness and distance. From this we can estimate
the total event rate for lenses bright enough and near enough to be
within reach of Gaia. Equation (\ref{equ1}) shows that the
error on the lens mass will be entirely dominated by the uncertainty
in the parallax; if we scale the Gaia errors linearly from the bright
end to the faint end (i.e. from 0.01 mas at $G=15$ mag to 0.1 mas at
$G=20$ mag) and consider events for which we will be able to obtain a
parallax error of 20 per cent, we find that we will obtain around 1
event every two years over the whole sky for quasars down to
$i=24$. If we allow for weakly magnified events (as above), this means
that LSST will be able to detect around one event per year for which
we will be able to robustly determine the distance (and hence lens
mass) to within 20 per cent accuracy using a parallax measurement from Gaia.

Another potential source of accurate parallax measurements is the
proposed SIM-lite space mission \citep{un08}, which, if it proceeds,
will enable significantly better mass determinations. With an
astrometric accuracy of 0.01 mas down to $V=20$ mag, this will be able
to follow-up a similar number of events to Gaia but with a ten-fold
improvement in the accuracy. Even though the precision 
of the mass recovery will be greatly improved, the total number of
events will not increase dramatically for SIM-lite since the limiting
magnitude will be similar to Gaia; the only advantage to the event
rate is the fact that the increased accuracy means it will be able to detect
parallaxes for more distant lenses, which results in a doubling of the
event rate compared to Gaia (i.e. around 2 events per year for LSST
using $i=24$ quasars).

Even if we are unable to obtain accurate trigonometric parallax
measurements, it will still be possible to accurately determine the
proper motion for a significant number of lenses using ground-based
astrometry. For example, LSST will be able to carry out astrometry
at around 3 mas for $r=24$ mag. As can be seen from the lens
luminosity function (Fig. \ref{fig:lens_luminosity_marg}) this covers
a significantly larger number of events, many of which will have
measurable proper motion (see Fig. \ref{fig:lens_propermotion_mag}). This
would almost entirely break the lens mass degeneracy, but to obtain a
unique mass measurement requires additional information, for example
the distance (from photometric/spectroscopic techniques) or through
higher-order microlensing effects. One potential
higher-order effect is microlensing parallax \citep{gou92}, where the
motion of the Earth as it orbits the Sun induces perturbations on the
standard microlensing signal. Simulations of parallax frequency show
that for distant sources the rate of parallax events towards the
Galactic bulge can reach around 10 per cent \citep[e.g. fig. 7
of][]{smi05}. This fraction will probably increase for lines-of-sight
away from the bulge, since the velocity dispersion of disc stars is
smaller than bulge stars and hence the event time-scale will increase
\citep[and hence the parallax fraction will also increase as this is related
to time-scale; e.g. see fig. 6 of][]{smi05}. As is shown in
Fig. \ref{fig:tE}, we find that 10 per cent of all events are
predicted to have time-scales greater than three months.

\begin{table}
\begin{tabular}{lc}
\hline
Lens Type & Event Fraction\\
& (per cent)\\
\hline
Brown dwarf & 7.5\\
Main sequence & 78.4\\
White dwarf & 13.1\\
Neutron star & 0.8\\
Black hole & 0.2\\
\hline
\end{tabular}
\caption{The fraction of different lenses by population, following our
simplistic implementation of stellar evolution (see Section
\ref{mod-mass-function}). These fractions are obtained trivially via
the IMF (equation \ref{eq:imf}).}
\label{tab:lens_population}
\end{table}

\section{Conclusion and discussion}
\label{sec:conclusion}

In this paper we have carried out Monte Carlo simulations to make
predictions for the microlensing of quasars by stars (and stellar
remnants) in the Milky Way. 
Using a simple model of the Galaxy we have predicted the all-sky
microlensing optical depth and have calculated the time-scale
distributions and event rates for future large-area sky surveys, both
for $i$- and $z$-band.
We found that the total event rate increases rapidly with increasing
magnitude limit, and predict that around 10 to 20 events will be
detected per year from surveys like Pan-STARRS and LSST (see Section
\ref{sec:event_rate}). Most events are caused by lenses in the thin
disc, even though extinction hampers the detectability of
low-latitude events. If we could identify and monitor large numbers of
quasars at low latitudes \citep[e.g.][]{lee08}, then it may even be
possible to increase the number of detected events.

We have also highlighted how such quasar microlensing can be used to carry out
robust and model-independent lens mass determinations, predicting that with
accurate astrometry from Gaia or SIM-lite it will be possible to
accurately recover the lens mass for around one event per
year. Microlensing is uniquely placed to provide mass measurements for
individual stars (as opposed to those in binary systems) and as such
will undoubtedly play a crucial role in stellar astronomy.

Although these results are very promising, there are a couple of
technical challenges which must be overcome in order to detect these
microlensing events. In this paper we have assumed a detection
efficiency of 100 per cent, which we have justified by noting that
practically all events should be reasonably well-sampled by surveys
like LSST. However, there are practical difficulties which will
potentially reduce the efficiency.

Firstly, it is well-known that quasars themselves are intrinsically
variable \citep[e.g.][]{koz10}. They can exhibit both 
long-term low-level variability as well as outburst phenomena. Indeed,
one recent paper claimed to have detected a spectacular UV flare in a
quasar behind M31 \citep{meu10}, although the authors admit that an
alternative explanation could be microlensing. Some authors have
claimed that nearly all quasars are microlensed by intervening
(cosmological) compact objects \citep[e.g.][]{haw96}, although more recent
studies have challenged this hypothesis \citep[e.g.][]{dev05}. This
latter paper analyses a variability diagnostic (the structure
function) of around 40,000 quasars, concluding that the most-likely
cause of quasar variability is flares due to disc instabilities. They
remark that the variability is, in general, both chromatic and
asymmetric. This is an important issue since the microlensing events
which we are considering in this paper should be both achromatic and
symmetric. Therefore one avenue for disentangling microlensing from
intrinsic variability would be to utilise the multi-band photometry
from surveys such as LSST. In addition, the amplitude of the quasar
variability is generally at a lower level than our microlensing
amplitude (e.g. \citealp{ses07} show that the rms scatter for quasars
is $\sim 0.07$ mag in the $g$-band and, because the rms decreases for
longer wavelengths, the variatbility will be even less for
microlensing surveys carried out in $i$- or $z$-bands).

Although most microlensing events are symmetric and achromatic, this
is not always the case. There are higher-order microlensing effects
which can cause deviations from the standard symmetric and chromatic
form, such as binarity in the lens or finite-source effects. Both of
these will probably be of only limited importance; for standard
galactic microlensing the fraction of events which show clear binary
signatures is of the order 3 per cent \citep[e.g.][and references
therein]{sko08}, while for finite-source effects it is even
less. Although it is safe to assume that the fraction of binary lens
events for quasar sources will be similar to that for bulge sources,
there is no reason to assume this will also hold for finite-source
effects. If we assume that a typical quasar accretion disc is of the
order $10^{13}$ cm, then the angular size will be less than a few
$\mu$as. This angular size is comparable to a giant star in the
Galactic bulge, which means that the finite-source effect for quasar
microlensing should be no-more important than for standard Galactic
microlensing; for these events finite-source effects only come into
play when the amplification reaches around one hundred or more
\citep[e.g.][]{cas06}. Therefore for the vast majority of quasar
microlensing events the light-curves will indeed be symmetric and
chromatic. For the limited number of events for which the
magnification is sufficient to detect finite-source signatures,
microlensing will enable constraints to be placed on the profiles of
quasar accretion discs, although this could only be done in very
fortuitous circumstances (e.g. when a low-redshift quasar is very
highly magnified).

A second obstacle to detecting microlensing of
quasars is the difficulty of identifying the quasars in the presence
of a luminous lens. The quasars will be identified via cuts in
colour-colour space, and so if the lens is sufficiently bright it
could compromise the quasar selection. Fortunately in most cases the
lenses will be significantly fainter than the sources (see
Fig. \ref{fig:lens_luminosity_marg}). We calculate that for only a
quarter of LSST events will the lens be brighter than the source and
for just over half of all events the lenses will be fainter than
the single-epoch magnitude limit of $i=24$.
However, this issue will be particularly important
for the events from which we wish to determine the lens mass, since
for these we $require$ the lens to be luminous so that its distance
and proper motion can be determined. For these events it will be
important to spectroscopically confirm the presence of a quasar behind
the lens, or to obtain high-resolution imaging a number of years after
the event so that the lens and quasar can be resolved.

In contrast to the above challenges that hamper the detection of quasar
microlensing events, there are avenues which could enhance our
ability to detect events. If we could build up an accurate astrometric
map of the foreground lenses and their proper motion (using, for
example, data from the Gaia mission) then it could be possible to
$predict$ which quasars will be lensed in advance of the
amplification. This method has been suggested as an approach to detect
lensing by pulsars \citep{dai10} but it is equally applicable to any
lens with an accurate proper motion measurment. Furthermore, this
could be applied to fainter quasars, such as those provided by the
complete co-added LSST survey, which will reach a depth of around
$i=27$. At this magnitude the number density of quasars climbs to
around 3000 per square degree (see Fig. \ref{fig2}.)
Another avenue is that of astrometric microlensing
where, instead of detecting the photometric lensing signatures, we
identify the astrometric shift of the image centroid as the quasar is
lensed \citep[e.g.][]{hog95,wal95}. This will be a demanding task as
it requires very high precision astrometry for our faint source
population; the maximum astrometric shift is 0.354 times the Einstein
radius \citep{dom00} and for our events typical Einstein radii are
around 1 mas. As above we could envisage targeting specific quasars
based on predictions from lens proper motions, but this would still
require astrometry at a level of at least tens of $\mu$as.

In this work we have only considered lenses within our own Galaxy, yet
one might wonder whether quasars could be lensed by stellar mass
objects at cosmological distances. However, this is a remote
possibility as it would require a very high density of such objects
and even then the time-scales for such events are a couple of orders
of magnitude longer than what we find for Galactic lenses \citep[see,
for example,][]{wam06}]. The notable exception to this, as noted in
the Introduction, is the phenomena of microlensing of 
macro-lensed (i.e. multiple-imaged) quasars by stellar mass lenses in
the intervening lensing galaxy. Here the density of lenses is
sufficient to produce an optical depth of order unity and hence
detecting microlensing becomes feasible. Surveys such as LSST and
Pan-STARRS will have an immense impact on this field as they will
monitor the light-curves for many thousands of multiply-imaged quasars.

Although we are unlikely to find events from lenses at cosmological
distances (with the exception of macro-lensed quasars), it may be
possible to detect lenses from populations just outside our own
Galaxy. For example, one could identify quasars behind the LMC
\citep[e.g.][]{koz09} or M31 \citep{huo10} and study their lensing by
foreground stars in these galaxies. As we mentioned above, there may
have already been an event of this kind detected in a quasar behind
M31 \citep{meu10}.

In conclusion, we have shown that imminent ground-based surveys will
be able to detect a number of quasar microlensing events every
year. The efforts of Pan-STARRS and LSST will, after five years of
operation, detect up to one hundred events. Although this is still too
few events to constrain Galactic models due to the overwhelming
numbers of parameters, it will allow for valuable consistency checks
to be carried out on these models and could lead to exciting and
unexpected discoveries.

\section*{Acknowledgments}

The authors are deeply indebted to Shude Mao for advice and guidance
throughout this work and to a helpful referee who enabled us to
improve the clarity of a number of issues.
JW is supported by the National Basic Research
Program of China (Grant No. 2009CB824800). MCS acknowledges support
from the Peking University One Hundred Talent Fund (985) and NSFC
grants 11043005 and 11010022 (International Young Scientist).


\begin{thebibliography}{}

\bibitem[Alcock et al.(1993)]{alc93} Alcock C. et al., 1993, Nature, 365, 621
\bibitem[Anguita et al.(2008)]{ang08} Anguita T., Faure C., Yonehara A., Wambsganss J., Kneib J.-P., Covone G., Alloin D., A\&A, 2008, 481, 615
\bibitem[Aubourg et al.(1993)]{aub93} Aubourg E. et al., 1993, Nature, 365, 623
\bibitem[Bailer-Jones (2009)]{bai09} Bailer-Jones C.A.L., 2009, in J.~Andersen, J.~Bland-Hawthorn, \& B.~Nordstr{\"o}m, eds, Proc. IAU Symp. 254, The Galaxy Disk in Cosmological Context, p. 210
\bibitem[Bastian, Covey \& Meyer (2010)]{bas10} Bastian N., Covey K.R., Meyer M.R., 2010, ARA\&A, 48, arXiv:1001.2965
\bibitem[Binney \& Tremaine (1987)]{bin87} Binney J., Tremaine S., 1987, Galactic Dynamics (Princeton; Princeton Univ. Press)
\bibitem[Binney \& Tremaine (2008)]{bin08} Binney J., Tremaine
  S., 2008, Galactic Dynamics (2nd ed.; Princeton Univ. Press)~(BT08)
\bibitem[Bond et al.(2001)]{bon01} Bond I. A. et al., 2001, MNRAS, 327, 868
\bibitem[Boyle et al. (2000)]{boy00} Boyle B. J. et al., 2000, MNRAS, 319, 1014
\bibitem[Byalko (1970)]{bya70} Byalko A. V., 1970, Soviet Astronomy, 13, 784
\bibitem[Cassan et al. (2006)]{cas06} Cassan A. et al., 2006, A\&A, 460, 277
\bibitem[Calchi Novati et al.(2005)]{cal05} Calchi Novati S. et al., 2005, A\&A, 443, 911
\bibitem[Chang \& Refsdal (1979)]{cha79} Chang K., Refsdal S., 1979, Nature, 282, 561
\bibitem[Coles et al (2009)]{col09} Coles J., Saha P., Schmid H. M., 2009, arXiv:0912.0515
\bibitem[Cox (1998)]{cox98} Cox A. N., 1999, Allen's Astrophysical Quantities, Fourth Edition, (Springer-Verlag: New York), 489
\bibitem[Croom et al. (2009)]{cro09} Croom S. M. et al., 2009, MNRAS, 399, 1755
\bibitem[Dominik \& Sahu (2000)]{dom00} Dominik M., Sahu, K.C., 2000, ApJ, 534, 213
\bibitem[Dai, Xu \& Esamdin (2010)]{dai10} Dai S., Xu R.X., Esamdin A., 2010, MNRAS, in press (arXiv:0912.1167)
\bibitem[de Jong et al.(2006)]{dej06} de Jong J. T. A. et al., 2006, A\&A, 446, 855
\bibitem[de Jong et al.(2010)]{dej10} {{de Jong} J.T.A., {Yanny} B.,
  {Rix} {H.-W.}, {Dolphin} A.E., {Martin} N.F., {Beers} T.C.}, ApJ,
  714, 663
\bibitem[de Vries et al. (2005)]{dev05} de Vries W.H., Becker R.H., White R.L., Loomis C., 2005, AJ, 129, 615
\bibitem[Einstein (1936)]{ein36} Einstein A., 1936, Sci, 84, 506
\bibitem[Fontanot et al. (2007)]{fon07} Fontanot F. et al., 2007, A\&A, 461, 39
\bibitem[Fukui et al.(2007)]{fuk07} Fukui A. et al., 2007, \apj, 670, 423
\bibitem[Gaudi et al.(2008)]{gau08} Gaudi B. S. et al., 2008, \apj, 677, 1268
\bibitem[G\'orski et al. (2005)]{gor05} G\'orski K. M. et al., 2005, \apj, 622, 759
\bibitem[Gould (1992)]{gou92} Gould A., 1992, \apj, 392, 442
\bibitem[Gould (2000)]{gou00} Gould A., 2000, \apj, 535, 928
\bibitem[Griest (1991)]{gri91} Griest K., 1991, \apj, 366, 412
\bibitem[Gwinn et al. (1997)]{gwi97} Gwinn C. R. et al., 1997, \apj, 485, 87
\bibitem[Hamadache et al.(2006)]{ham06} Hamadache C. et al., 2006, A\&A, 454, 185
\bibitem[Han (2008)]{han08} Han C., 2008, \apj, 681, 806
\bibitem[Han \& Gould (1995)]{han95} Han C., Gould A., 1995, \apj, 447, 53
\bibitem[Hawkins (1996)]{haw96} Hawkins M.R.S., 1996, MNRAS, 278 787
\bibitem[Hennawi et al. (2007)]{hen07} Hennawi J. F., Dalal N., Bode P., 2007, \apj, 654, 93
\bibitem[H{\o}g, Novikov \& Polnarev (1995)]{hog95} H{\o}g E., Novikov I.D., Polnarev A.G., 1995, A\&A, 294, 287
\bibitem[Huo et al. (2010)]{huo10} Huo Z.-Y. et al., 2010, RAA, 10, 612
\bibitem[Irwin et al. (1989)]{irw89} Irwin M. J. et al., 1989, \aj, 98, 1989
\bibitem[Ivezi{\'c} et al. (2008)]{ive08} Ivezi{\'c} {\v Z}., Tyson J.~A., Allsman R., Andrew J., Angel R., 2008, arXiv:0805.2366
\bibitem[Jordi et al (2006)]{jor06} Jordi K., Grebel E. K., Ammon K., 2006, A\&A, 460, 339
\bibitem[Kaiser et al. (2002)]{kai02} Kaiser N. et al., 2002, SPIE, 4836, 154
\bibitem[Kayser, Refsdal \& Stabell (1986)]{kay86} Kayser R., Refsdal S., Stabell R., 1986, A\&A, 166, 36
\bibitem[Kiraga \& Paczy\'nski (1994)]{kir94} Kiraga M., Paczy\'nski B., 1994, \apj, 430, L101
\bibitem[Koz{\l}owski \& Kochanek (2009)]{koz09} Koz{\l}owski S., Kochanek C. S., 2009, ApJ, 701, 508
\bibitem[Koz{\l}owski et al. (2010)]{koz10} Koz{\l}owski S. et al., 2010, ApJ, 708, 927
\bibitem[Kroupa (2002)]{kro02} Kroupa P., 2002, Sci, 295, 82
\bibitem[Lee et al. (2008)]{lee08} Lee I. et al., 2008, ApJS, 175, 116
\bibitem[Li et al. (2007)]{lig07} Li G. L. et al., 2007, MNRAS, 378, 469
\bibitem[Liebes (1964)]{lie64} Liebes S., 1964, Phys. Rev., 133, 835
\bibitem[Mao (2008)]{mao08} Mao S., 2008, arXiv:0811.0441
\bibitem[Meusinger et al. (2010)]{meu10} Meusinger H., 2010, A\&A, 512, 1
\bibitem[Paczy\'nski (1986)]{pac86} Paczy\'nski B., 1986, \apj, 304, 1
\bibitem[Paczy\'nski (1991, 1996)]{pac91} Paczy\'nski B., 1991, \apj, 371, L63
\bibitem[Paczy\'nski (1995)]{pac95} Paczy\'nski B., 1995, Acta Astronomica, 45, 345
\bibitem[Paczy\'nski (1996)]{pac96} Paczy\'nski B., 1996, ARA\&A, 34, 419
\bibitem[Palanque-Delabrouille et al.(1998)]{pal98} Palanque-Delabrouille N. et al., 1998, A\&A, 332, 1
\bibitem[Paulin-Henriksson et al.(2002)]{pau02} Paulin-Henriksson S. et al., 2002, \apj, 576, L121
\bibitem[Perryman et al.(2001)]{per01} Perryman M.A.C. et al., 2001, A\&A, 369, 339
\bibitem[Popowski et al. (2005)]{pop05} Popowski P. et al., 2005, ApJ, 631, 879  
\bibitem[Refsdal (1964)]{ref64} Refsdal S., 1964, MNRAS, 128, 295
\bibitem[Renn, Sauer \& Stachel (1997)]{ren97} Renn J., Sauer T., Stachel J., 1997, Sci, 275, 184
\bibitem[Richards et al. (2006)]{ric06} Richards G. et al., 2006, \aj, 131, 2766
\bibitem[Richards et al. (2009)]{ric09} Richards G. et al., 2009, \apj, 180, 67
\bibitem[Sesar et al. (2007)]{ses07} Sesar B. et al., 2007, AJ, 134, 2236
\bibitem[Schlegel et al.(1998)]{sch98} Schlegel D., Finkbeiner D. P., and Davis M., 1998, \apj, 500, 525~(SFD)
\bibitem[Sch{\"o}nrich, Binney \& Dehnen (2010)]{sch10} {{Sch{\"o}nrich} R., {Binney} J., {Dehnen} W.}, MNRAS, in press (arXiv:0912.3693)
\bibitem[Skowron et al. (2008))]{sko08} Skowron J. et al., 2008, Acta Astron., 57, 281
\bibitem[Smith et al. (2005))]{smi05} Smith M. C., Belokurov V., Evans N.W., Mao S., An J.H., 2005, MNRAS, 361, 128
\bibitem[Sumi et al. (2006))]{sum06} Sumi T. et al., 2006, ApJ, 636, 240
\bibitem[Smith et al. (2007))]{smi07} Smith M. C. et al., 2007, MNRAS, 379, 755
\bibitem[Tisserand et al.(2007)]{tis07} Tisserand P. et al., 2007, A\&A, 469, 387
\bibitem[Thomas et al.(2005)]{tho05} Thomas C. L. et al., 2005, \apj, 631, 906
\bibitem[Tyson (2002)]{tys02} Tyson J.A., 2002, SPIE, 4836, 10
\bibitem[Udalski et al.(1992)]{uda92} Udalski A. et al., 1992, Acta Astron., 42, 253
\bibitem[Udalski et al.(2000)]{uda00} Udalski A. et al., 2000, Acta Astron., 50, 1
\bibitem[Unwin et al.(2008)]{un08} Unwin S.C. et al., 2008, PASP, 120, 38
\bibitem[Uglesich et al.(2004)]{ugl04} Uglesich R. R., 2004, \apj, 612, 877
\bibitem[Walker (1995)]{wal95} Walker M.A., 1995, ApJ, 453, 37
\bibitem[Wambsganss (2006)]{wam06} Wambsganss J., 2006, in G.~Meylan,
P.~Jetzer, P.~North, P.~Schneider, C.~S.~Kochanek, \& J.~Wambsganss,
eds, Saas-Fee Advanced Course 33: Gravitational Lensing: Strong, Weak and Micro
\bibitem[Wood \& Mao (2005)]{woo05} Wood A., Mao S., 2005, MNRAS, 362, 945
\bibitem[Wyrzykowski et al. (2009)]{wyr09} Wyrzykowski {\L}. et al., 2009, MNRAS, 397, 1228
\bibitem[Wyrzykowski et al. (2010)]{wyr10} Wyrzykowski {\L}. et al., 2010, MNRAS, in press (arXiv:1004.5247)
\end{thebibliography}
\end{document}